\documentclass[preprint,12pt,number]{elsarticle}
\usepackage[T1]{fontenc}
\usepackage{lmodern}
\usepackage{amsmath,amssymb}
\usepackage{graphicx}
\usepackage{booktabs}
\usepackage{multirow}
\usepackage{array}
\usepackage{tabularx}
\usepackage{adjustbox}
\usepackage{microtype}
\usepackage{url}
\usepackage{xcolor}
\usepackage{tikz}
\usetikzlibrary{arrows.meta,positioning,calc,fit,backgrounds,shapes.geometric,shadows.blur}
\usepackage[hidelinks]{hyperref}
\hypersetup{pdfauthor={Amir Mohammad Khadem Hosseini and Sattar Mirzakuchaki}}
\journal{Computers \& Electrical Engineering}
\biboptions{numbers,sort&compress}

\newcommand{\NumParams}{149,835}
\newcommand{\ExecParams}{149,423}
\newcommand{\SynWeights}{143,824}

\newcommand{\NumPLIF}{30,976,000}
\newcommand{\NumAdapt}{10,982,400}

\definecolor{ink}{HTML}{302B2B}
\definecolor{muted}{HTML}{625D5D}
\definecolor{paper}{HTML}{FFFDFC}
\definecolor{preDark}{HTML}{5C8FBC}
\definecolor{preMid}{HTML}{87ADD0}
\definecolor{preLight}{HTML}{EEF5FB}
\definecolor{encDark}{HTML}{5E966B}
\definecolor{encMid}{HTML}{8DBA91}
\definecolor{encLight}{HTML}{F1F8EF}
\definecolor{bridgeDark}{HTML}{7D4F9F}
\definecolor{bridgeMid}{HTML}{A47BC0}
\definecolor{bridgeLight}{HTML}{F5EFF9}
\definecolor{decDark}{HTML}{B73544}
\definecolor{decMid}{HTML}{D76A73}
\definecolor{decLight}{HTML}{FCEFF0}
\definecolor{readDark}{HTML}{263F79}
\definecolor{readMid}{HTML}{5B73A4}
\definecolor{readLight}{HTML}{F0F3FA}
\definecolor{plifDark}{HTML}{B66B2F}
\definecolor{plifLight}{HTML}{FCE7D2}
\definecolor{neutralLight}{HTML}{F4F2F1}
\definecolor{neutralEdge}{HTML}{7D7774}

\newcommand{\FigureHeaderFont}{\fontsize{9.7}{11.0}\selectfont\bfseries}

\newcommand{\FigureBodyFont}{\fontsize{7.9}{9.2}\selectfont}
\newcommand{\FigureBodyBold}{\fontsize{7.9}{9.2}\selectfont\bfseries}
\newcommand{\FigureTensorFont}{\fontsize{7.6}{8.9}\selectfont\bfseries}

\newcommand{\FigureNoteFont}{\fontsize{7.1}{8.3}\selectfont}
\newcommand{\SJConv}{\textbf{SpikingJelly Conv2d}}
\newcommand{\PLIFNode}{\textbf{ParametricLIF}}

\tikzset{
  flow/.style={-{Latex[length=3.0mm,width=2.1mm]}, line width=1.15pt, draw=ink, rounded corners=2pt},
  skip/.style={-{Latex[length=2.7mm,width=1.9mm]}, line width=1.0pt, draw=ink!88, dashed, dash pattern=on 4pt off 3pt, rounded corners=4pt},
  tensor/.style={font=\FigureTensorFont, text=ink, align=center, fill=white, draw=ink!18, rounded corners=3pt, inner xsep=4pt, inner ysep=3pt},
  note/.style={font=\FigureNoteFont, align=center, text=muted},
  card/.style={draw=ink!24, fill=white, rounded corners=4pt, line width=0.95pt, align=center, font=\FigureBodyFont, minimum height=9.5mm, minimum width=21mm, inner xsep=4pt, inner ysep=3pt},
  precard/.style={card, draw=preDark!70},
  enccard/.style={card, draw=encDark!78},
  bridgecard/.style={card, draw=bridgeDark!78},
  deccard/.style={card, draw=decDark!80},
  readcard/.style={card, draw=readDark!80},
  neutralcard/.style={card, draw=neutralEdge!65, fill=neutralLight!45},
  plif/.style={draw=plifDark!90, fill=plifLight, rounded corners=5pt, line width=1.0pt, minimum width=24mm, minimum height=9.5mm, align=center, font=\FigureBodyBold},
  add/.style={circle, draw=ink!72, fill=white, line width=1.25pt, minimum size=6mm, inner sep=0pt, font=\bfseries\small}
}
\newcommand{\panelbg}[4]{%
  \begin{scope}[on background layer]
    \node[fit=#1, rounded corners=8pt, fill=#2, draw=#3, line width=1.15pt, inner sep=8pt] (#4) {};
  \end{scope}}
\newcommand{\panelheader}[4]{%
  \node[fill=#2, rounded corners=4pt, minimum width=#4, minimum height=8mm, anchor=south west] (#1hdr) at ($(#1.north west)+(0mm,1.5mm)$) {};
  \node[font=\FigureHeaderFont,text=white,anchor=west] at ($(#1hdr.west)+(4mm,0)$) {#3};}

\begin{document}
\begin{frontmatter}
\title{EventSpikeSeg: Memory-Aware Mixed-Signal Architecture for Event-Camera Spiking Segmentation}
\author[iust]{Amir Mohammad Khadem Hosseini\corref{cor1}}
\ead{am_khadem@alumni.iust.ac.ir}
\author[iust]{Sattar Mirzakuchaki}
\ead{m_kuchaki@iust.ac.ir}
\cortext[cor1]{Corresponding author.}
\address[iust]{Iran University of Science and Technology, Hengam Street, Resalat Square, Tehran 16846-13114, Iran}
\begin{abstract}
Event-camera segmentation requires persistent neural state despite sparse input events, making memory organization a central constraint on embedded implementation. We present EventSpikeSeg, a mixed-signal architecture combining resident capacitor memory with shared analog processing in GF180MCU 180-nm CMOS. Its 25-population spiking encoder--decoder has 143,824 synaptic weights but requires 30,976,000 membrane states at $440\times640$ resolution. The mapping specifies chronological 5-ms model updates, digital event/address services and 64-state capacitor banks. Characterization covers 22 module families and 74 operating points, distinguishing transistor simulations from digital-library estimates. Nominal results include 41.711-$\mu$W biased neuron-core power, 3.338-pJ positive terminal energy for a 1.4-$\mu$s cell read, and 0.531-mV selected-state drift over a 5-ms extracted-tile hold. Commit and coupling errors are reported separately. The software checkpoint records 27.976\% mean intersection-over-union on a held-out six-class DSEC split; this is not hardware accuracy. Component layouts and simulation evidence quantify local implementation costs and expose storage replication and service scheduling as scaling constraints. A detailed SPICE-calibrated behavioral forward simulation of one full-resolution DSEC frame achieves 24.26\% mean intersection-over-union, versus 26.22\% for the matched PyTorch reference, using a compact 5-fF bank variant and external count/voltage calibration. Physical full-network operation, system energy and mismatch robustness remain unverified.
\end{abstract}
\end{frontmatter}

\section{Introduction}
Event cameras emit asynchronous brightness-change events $e=(x,y,t,p)$ rather than conventional intensity frames. Their high temporal resolution and activity-dependent output motivate embedded perception pipelines that do not discard timing information at the sensor boundary \cite{gallego2022survey,binas2017ddd17,gehrig2021dsec}. Directly trained spiking neural networks (SNNs) can retain time-dependent internal state while communicating hidden activations as discrete spikes \cite{neftci2019surrogate,fang2021plif,fang2023spikingjelly}. Semantic segmentation nevertheless creates a particularly demanding architectural workload: dense per-pixel class decisions require both long-range context and native-resolution detail, even when activity at the input is sparse.

A small number of trainable synaptic weights is not an adequate predictor of the area, memory bandwidth, or response time of a spatially expanded recurrent SNN. Every active pixel/channel instance can retain a membrane variable, and additional adaptation and readout variables persist between model updates. A systems-oriented evaluation must therefore account for the software execution graph, event routing and accumulation, resident state placement, shared compute service, update ordering, and finally the coverage of physical evidence. Circuit-level power numbers alone cannot establish system-level efficiency.

This article connects event-camera computation to electrical implementation through EventSpikeSeg, examining how persistent analog storage and shared service circuitry realize a stateful inference workload. The workload is a six-class event-camera encoder--decoder with persistent PLIF state and a separate native-resolution detail branch. Its research implementation uses GF180MCU analog storage, shared update/readout functions, and categorical digital dispatch. Native circuit layouts and a hierarchical research assembly provide the physical mapping evidence; the assembly is not a compact taped-out die. The electrical evidence consists of characterized modules and a connected selected-state experiment, complemented by a full-frame, SPICE-calibrated behavioral network simulation. Autonomous physical full-network execution remains unverified.

The contributions of the present study are as follows:
\begin{enumerate}
\item A workload-to-hardware mapping of a 25-actor, 54-edge spiking segmentation graph that explicitly distinguishes synaptic parameters from more than 54 million recurrent and readout-state identities.
\item An execution contract for chronological 5-ms updates, including signed event contributions, persistent-state restoration, sharing of analog update circuits, and the boundary between model-time requirements and unverified full-population deadlines.
\item A banked mixed-signal implementation with native component layouts and 74 reported operating points across 22 module families, including an extracted-tile signed-update and retention experiment.
\item An evaluation connecting module simulations to a detailed calibrated full-frame forward pass, with a matched software reference, storage-area estimates, and explicit system-level limits.
\end{enumerate}

The aim is to document the architectural consequences of applying a stateful dense-prediction model to neuromorphic hardware; the results do not demonstrate a fabricated device, end-to-end hardware accuracy, or lower system energy than published processors. Sections~\ref{sec:model} and~\ref{sec:mapping} specify the workload and mapping, Sections~\ref{sec:asic} and~\ref{sec:verification} cover implementation, and Section~\ref{sec:results} distinguishes the completed evidence from open validation questions.

\section{Related Work}
\subsection{Event-Based Semantic Segmentation}
DSEC and DSEC-Semantic provide high-resolution driving events and semantic labels suitable for raw-event dense prediction \cite{gehrig2021dsec,sun2022ess}. EV-SegNet established an early event-only semantic-segmentation baseline \cite{alonso2019evsegnet}. Kim \emph{et al.} demonstrated direct surrogate-gradient training for semantic segmentation \cite{kim2022beyond}; EvSegSNN emphasized a compact PLIF U-shaped network \cite{hareb2024evsegsnn}. More recent SpikingEDN and SLTNet designs introduced adaptive thresholds, multi-path processing, and spike-driven attention-like operations \cite{zhang2024spikingedn,long2025sltnet}.

\subsection{Neuromorphic and Hardware-Oriented SNN Processing}
Loihi-based event processing and hardware-aware pipelines such as TRIP illustrate the benefit of aligning event sensing, SNN dynamics, and hardware execution \cite{renner2019attention,arjmand2024trip}. FPGA-oriented work similarly emphasizes fixed-point neuron state and explicit event routing \cite{courtois2024embedded}. Here, the model graph is mapped to a fixed application-specific directed acyclic graph (DAG) of actors, with explicit accounting for every spatially replicated recurrent state.

Patel \emph{et al.} provide a related ASIC deployment: a spiking U-Net mapped across Intel Loihi chips for microscopy-image segmentation \cite{patel2021segmentation}. Using 8-bit weights across platforms, they report over twice the energy efficiency of CPU/GPU execution for online single-image inference, with comparable task performance. Their work demonstrates dense prediction on digital neuromorphic silicon and addresses inter-chip communication. EventSpikeSeg instead maps event-camera segmentation to analog membrane storage and shared mixed-signal circuitry. The different datasets and measured-chip versus module-simulation evidence preclude a direct efficiency ranking.

Mixed-signal processors provide a closer circuit-level context. DYNAPs combines analog neuron/synapse dynamics with asynchronous digital routing and heterogeneous memory structures, demonstrated in prototype silicon \cite{moradi2018dynaps}. Braindrop couples subthreshold analog dynamics to a high-level programming model and compensates device variability at network level \cite{neckar2019braindrop}. EventSpikeSeg investigates a complementary mapping problem: retaining the spatial state inventory of one trained segmentation graph while sharing analog service circuitry. Its evidence comprises hierarchical layout, nominal module simulation, and a calibrated behavioral forward pass on one complete event-camera record. These different workloads, implementation stages, and energy boundaries do not support a numerical energy-efficiency ranking against those chips.

\subsection{Embedded SNN Architecture and Co-design}
The Darwin neural processing unit illustrates a configurable digital SNN co-processor, with prototype-silicon evidence for an application-facing architecture \cite{ma2017darwin}. Recent work in this journal has also treated SNN placement and spike-communication patterns as a joint software/hardware optimization problem \cite{xiao2025codesign}. These studies establish an architectural perspective in which communication, memory, mapping, and scheduling matter alongside neuron arithmetic. EventSpikeSeg targets a different workload and a different evidence stage: dense event-based segmentation mapped to resident analog state with shared compute peripherals. Neither throughput nor energy rankings are inferred across these non-equivalent experiments. The present contribution is principally an explicit state-and-service mapping and its physical feasibility constraints, rather than a measured replacement for these systems.

\subsection{Open Mixed-Signal Implementation Flow}
The GF180MCU open PDK supplies device and physical-design resources for a 0.18-$\mu$m MCU process \cite{gf180pdk}. Native analog cells are assembled with digital control macros; Yosys and OpenROAD support the latter's RTL-to-layout flow \cite{ajayi2019openroad}. KLayout renders the delivered mask geometry directly from GDS \cite{klayout}. Earlier ASAP7 digital results and analog tool-compatibility probes \cite{clark2016asap7,vashishtha2017asap7} remain historical evidence, not the technology basis of the present analog results.

\section{Workload and Stateful Execution Model}
\label{sec:model}

The model specification defines the temporal and spatial requirements of the circuit implementation. EventSpikeSeg is the primary model and hardware target; EventSpikeSegXL is the larger capacity/reference model. The ten 5-ms bins define model time; achieving that schedule at full population is a separate hardware question.

\subsection{Raw-Event Protocol}
For semantic label time $t_k$, the DSEC loader selects rectified events
\begin{equation}
\mathcal{E}_k=\{e_i\mid t_k-50\,\mathrm{ms}\le t_i<t_k\}
\end{equation}
and assigns them chronologically to ten model updates of $\Delta t=5$ ms. The six active output classes merge the eleven DSEC semantic classes as: road/sidewalk $\rightarrow$ flat; sky/building/fence/wall $\rightarrow$ construction+sky; pole/traffic-sign $\rightarrow$ object; vegetation $\rightarrow$ nature; person $\rightarrow$ human; and vehicle $\rightarrow$ vehicle. Complete sequences are held out for validation so recurrent state is never evaluated across shuffled frames.

\subsection{PLIF and Custom Dynamics}
All signed synaptic contributions for one destination and one 5-ms interval are accumulated before one PLIF update. A simplified floating-point form is
\begin{align}
 u_k &= \beta v_{k-1}+I_k,\\
 s_k &= H(u_k-\theta_k),\\
 v_k &= u_k-s_k\theta_k.
\end{align}
Here $I_k$ is the accumulated synaptic contribution, $\beta$ is learned retention, $\theta_k$ is the effective threshold including adaptation, and $H$ is the threshold function. The subtractive reset follows the original float path with no fixed reset voltage. The model retains stem-only peer-membrane inhibition and triangular retinotopic threshold adaptation. Hidden winner-peer WTA is excluded because the mechanism screen showed severe deep-layer spike starvation.

\subsection{Inference Graph and Recurrent-State Inventory}
\begin{figure}[!t]
\centering
\begin{adjustbox}{max width=0.98\textwidth}
\begin{tikzpicture}[x=1cm,y=1cm,>=Latex,
 vbox/.style={rounded corners=3pt,line width=0.8pt,fill=white,draw=ink!35,text=ink,align=center,text width=3.65cm,minimum height=1.45cm,font=\fontsize{9.5}{11.5}\selectfont,inner sep=5pt},
 vf/.style={-{Latex[length=2.6mm]},draw=ink,line width=1.05pt,rounded corners=3pt},
 vb/.style={circle,fill=ink,text=white,font=\bfseries\small,inner sep=2.5pt}]
\begin{scope}[on background layer]
\filldraw[fill=preLight,draw=preMid,rounded corners=5pt] (-2.2,-1.05) rectangle (2.2,11.15);
\filldraw[fill=encLight,draw=encMid,rounded corners=5pt] (3.3,-1.05) rectangle (7.7,11.15);
\filldraw[fill=bridgeLight,draw=bridgeMid,rounded corners=5pt] (8.8,-1.05) rectangle (13.2,11.15);
\filldraw[fill=decLight,draw=decMid,rounded corners=5pt] (14.3,-1.05) rectangle (18.7,3.8);
\filldraw[fill=readLight,draw=readMid,rounded corners=5pt] (14.3,4.35) rectangle (18.7,11.15);
\end{scope}
\foreach \x/\col/\txt in {0/preDark/PREPROCESSING,5.5/encDark/ENCODER,11/bridgeDark/BRIDGE AND FUSION,16.5/readDark/READOUT} {
\node[fill=\col,text=white,rounded corners=3pt,minimum width=4.4cm,minimum height=0.55cm,font=\bfseries\small] at (\x,11.55) {\txt};}
\node[fill=decDark,text=white,rounded corners=3pt,minimum width=4.4cm,minimum height=0.45cm,font=\bfseries\small] at (16.5,4.05) {DECODER};
\node[vbox,draw=preDark] (raw) at (0,0) {\textbf{DSEC input}\\ raw events $(t,x,y,p)$};
\node[vbox,draw=preDark] (tensor) at (0,4.2) {\textbf{Event tensor}\\ 2 polarities; $440\times640$\\ 10 bins of 5 ms};
\node[vbox,draw=preDark] (stem) at (0,8.4) {\textbf{Spiking stem}\\ $5\times5$, stride 2\\ $2\to32$};
\node[vbox,draw=encDark] (e1) at (5.5,8.4) {\textbf{Encoder 1}\\ 32 channels\\ $220\times320$};
\node[vbox,draw=encDark] (e2) at (5.5,4.2) {\textbf{Encoder 2}\\ 64 channels\\ $110\times160$};
\node[vbox,draw=encDark] (e3) at (5.5,0) {\textbf{Encoder 3}\\ 128 channels\\ $55\times80$};
\node[vbox,draw=bridgeDark] (detail) at (11,8.4) {\textbf{Native detail path}\\ $2\to8\to8$\\ $440\times640$};
\node[vbox,draw=bridgeDark] (fuse) at (11,5.6) {\textbf{PLIF fusion}\\ $3\times3$, $40\to16$\\ $440\times640$};
\node[vbox,draw=bridgeDark] (cat) at (11,2.8) {\textbf{Native concatenation}\\ upsampled semantic32\\ + detail8: 40 channels};
\node[vbox,draw=bridgeDark] (ctx) at (11,0) {\textbf{Semantic context}\\ 128 channels; $55\times80$\\ two dilated residual blocks};
\node[vbox,draw=decDark] (d1) at (16.5,0) {\textbf{Decoder 1}\\ project $128\to64$\\ upsample; concat 64-channel skip\\ fuse $128\to64$};
\node[vbox,draw=decDark] (d2) at (16.5,2.8) {\textbf{Decoder 2}\\ project $64\to32$; upsample\\ add 32-channel skip};
\node[vbox,draw=readDark] (head) at (16.5,5.6) {\textbf{Continuous class head}\\ $1\times1$, $16\to6$\\ no output PLIF};
\node[vbox,draw=readDark] (integ) at (16.5,7.9) {\textbf{Temporal integration}\\ 10-bin integration\\ first-index argmax};
\node[vbox,draw=readDark] (mask) at (16.5,10.1) {\textbf{Segmentation mask}\\ 6 classes};
\draw[vf] (raw)--(tensor);\draw[vf] (tensor)--(stem);\draw[vf] (stem)--(e1);
\draw[vf] (e1)--(e2);\draw[vf] (e2)--(e3);\draw[vf] (e3)--(ctx);\draw[vf] (ctx)--(d1);
\draw[vf] (d1)--(d2);\draw[vf] (d2)--(cat);\draw[vf] (cat)--(fuse);\draw[vf] (fuse)--(head);\draw[vf] (head)--(integ);\draw[vf] (integ)--(mask);
\draw[vf,draw=bridgeDark] (tensor.west)--(-2.5,4.2)--(-2.5,10.25)--(11,10.25)--(detail.north);
\node[font=\small,text=bridgeDark,fill=white,inner sep=2pt] at (5.0,10.25) {native input branch};
\draw[vf,draw=bridgeDark] (detail.west)--(8.45,8.4)--(8.45,2.8)--(cat.west);
\node[vb] at (7.35,8.4) {A};\node[vb] at (18.35,2.8) {A};
\node[vb] at (7.35,4.2) {B};\node[vb] at (18.35,0) {B};
\node[font=\small,text=ink,align=center] at (8.1,-1.6) {\textbf{A:} 32-channel skip sum\qquad\textbf{B:} 64-channel skip concatenation\\25 hidden PLIF actors; native detail and multiscale semantic path};
\end{tikzpicture}\end{adjustbox}
\caption{EventSpikeSeg inference topology used for the analog mapping. The native detail stream branches directly from the input. Matching A markers denote the half-resolution 32-channel skip sum; B markers denote the quarter-resolution 64-channel skip concatenation. Native fusion maps 40 to 16 channels, followed by a continuous 16-to-6 class-current head. The 25 hidden PLIF actors are enumerated in Table~\ref{tab:140k_layers}; auxiliary training heads are omitted. This is an architecture diagram, not a physical-layout image. Diagram preparation used AI-assisted LaTeX/TikZ coding, reviewed by the authors.}
\label{fig:base_arch}
\end{figure}
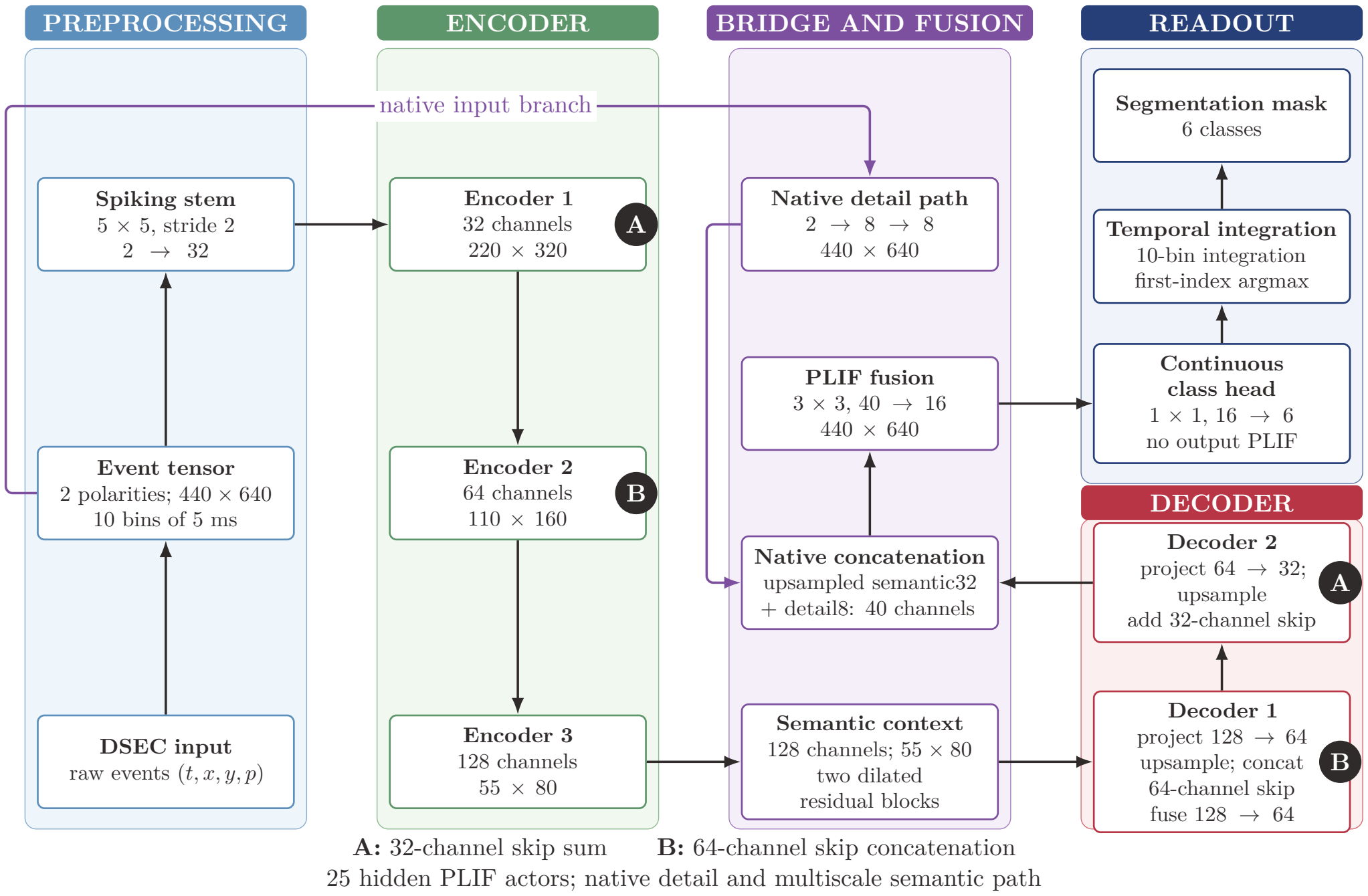

Fig.~\ref{fig:base_arch} shows the final model topology and its two skip merges.

The primary model has \NumParams{} registered parameters, \ExecParams{} inference parameters, and \SynWeights{} deployed synaptic weights. Its 25 PLIF populations form the exported 54-edge actor graph. Table~\ref{tab:140k_layers} expands the native detail, semantic encoder, context, and decoder paths. Actor counts include both PLIF stages of each depthwise/pointwise block; a pointwise projection with one PLIF contributes one actor.

\begin{table}[!t]
\centering\small
\caption{Inference topology of EventSpikeSeg at $440\times640$ input resolution. DW/PW denotes depthwise $3\times3$ followed by pointwise $1\times1$; each has its own affine transform and PLIF. Membrane counts exclude adaptation and continuous-readout state.}
\label{tab:140k_layers}
\begin{tabularx}{\textwidth}{p{2.1cm}p{2.0cm}X r r}
\toprule
Path / stage & Spatial size & Executed operations & Actors & Membranes \\
\midrule
Native detail & $440\times640$ & Standard $3\times3$: $2\to8$, then residual $8\to8$. & 2 & 4,505,600 \\
Half encoder & $220\times320$ & Standard $5\times5$, stride 2: $2\to32$; residual standard $3\times3$ and residual DW/PW at 32 channels. & 4 & 9,011,200 \\
Quarter encoder & $110\times160$ & Standard $3\times3$, stride 2: $32\to64$; DW/PW at 64 channels, then residual DW/PW. & 5 & 5,632,000 \\
Eighth encoder & $55\times80$ & DW/PW stride 2: $64\to128$; DW/PW at 128 channels, then residual DW/PW. & 6 & 3,097,600 \\
Dilated context & $55\times80$ & Two residual DW/PW blocks at 128 channels; depthwise dilation 2. & 4 & 2,252,800 \\
Decoder projections & Eighth / quarter & $1\times1$: $128\to64$ at eighth; upsample and concatenate with quarter skip ($64+64$); fuse $128\to64$, then project $64\to32$. & 3 & 1,971,200 \\
Native fusion & $440\times640$ & Upsample semantic32, merge half-resolution skip, upsample to native; concatenate detail8, standard $3\times3$ $40\to16$. & 1 & 4,505,600 \\
\midrule
Total hidden state & & & 25 & 30,976,000 \\
Continuous head & $440\times640$ & $1\times1$ signed current: $16\to6$, temporal accumulation; no class PLIF. & 0 & --- \\
\bottomrule
\end{tabularx}
\end{table}

\begin{figure}[!t]
\centering
\begin{adjustbox}{max width=0.96\textwidth}
\begin{tikzpicture}[node distance=6mm]
  \node[tensor] (in) {input\\$[T,N,C,H,W]$};
  \node[enccard, right=8mm of in, minimum width=31mm] (dw) {\SJConv\\depthwise, $3\!\times\!3$, $s=1$\\groups $=C$};
  \node[enccard, right=of dw, minimum width=24mm] (a1) {\textbf{ChannelAffine}};
  \node[plif, right=of a1] (p1) {\PLIFNode};
  \draw[flow] (in)--(dw)--(a1)--(p1);
  \node[enccard, below=15mm of dw, minimum width=33mm] (pw) {\SJConv\\pointwise, $1\!\times\!1$\\$C\!\rightarrow\!C$};
  \node[enccard, right=of pw, minimum width=24mm] (a2) {\textbf{ChannelAffine}};
  \node[plif, right=of a2] (p2) {\PLIFNode};
  \node[add, right=8mm of p2] (add) {$+$};
  \node[neutralcard, right=7mm of add, minimum width=18mm] (clamp) {clamp\\$[0,1]$};
  \node[tensor, right=8mm of clamp] (out) {output\\$[T,N,C,H,W]$};
  \draw[flow] (p1.south) -- ++(0,-6mm) -| (pw.north);
  \draw[flow] (pw)--(a2)--(p2)--(add)--(clamp)--(out);
  \draw[skip] (in.south) -- ++(0,-31mm) -| (add.south);
  \node[note, below=21mm of a2, xshift=27mm] (n) {identity residual is added \textbf{after the second ParametricLIF}};
  \panelbg{(in)(p1)(pw)(out)(n)}{encLight}{encDark!72}{panelB}
  \panelheader{panelB}{encDark}{RESIDUAL SPIKINGCONVBLOCK}{158mm}
\end{tikzpicture}
\end{adjustbox}
\caption{Depthwise/pointwise residual variant of \texttt{SpikingConvBlock}. Standard-convolution stages instead use one convolution, channel affine transform, and PLIF. The residual is used only for equal channel counts and stride one. Diagram preparation used AI-assisted LaTeX/TikZ coding, reviewed by the authors.}
\label{fig:resblock}
\end{figure}
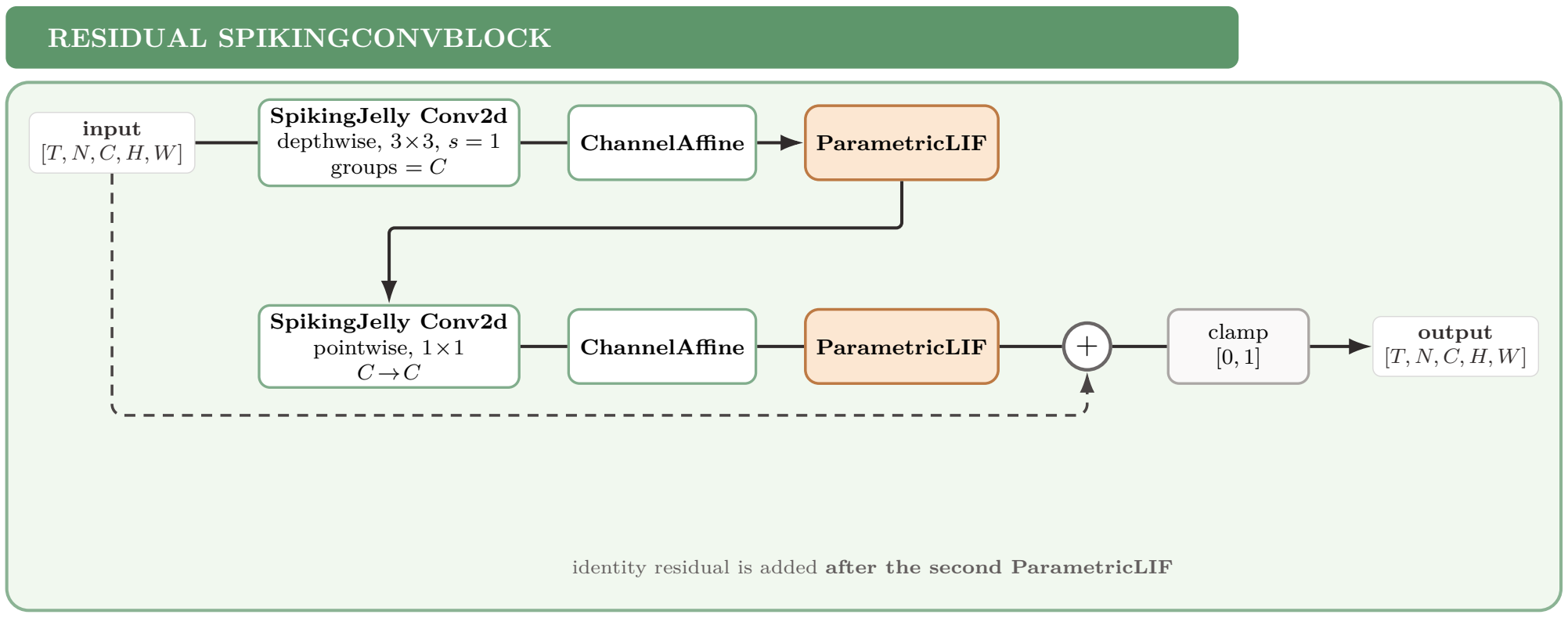

The half-resolution decoder skip adds semantic and encoder spikes while preserving multiplicity up to two; it is not clamped before native fusion. Within a residual feature block, the residual sum after PLIF is clamped to $[0,1]$. Fig.~\ref{fig:resblock} shows the depthwise and pointwise variant. Standard blocks use a single convolution--affine--PLIF stage. The context dilation enlarges the receptive field without widening the 128-channel state tensor.

The model's learnable retention and thresholds coexist with channel-affine transforms. Stem-only peer-membrane inhibition and encoder/context triangular threshold adaptation are retained; native detail and the half-resolution encoder path do not receive that adaptation. The exported inventory contains \NumAdapt{} adaptation-enabled identities. Auxiliary and boundary heads support training and are excluded from the executed inference inventory.

\subsection{Native Fusion and Continuous Readout}
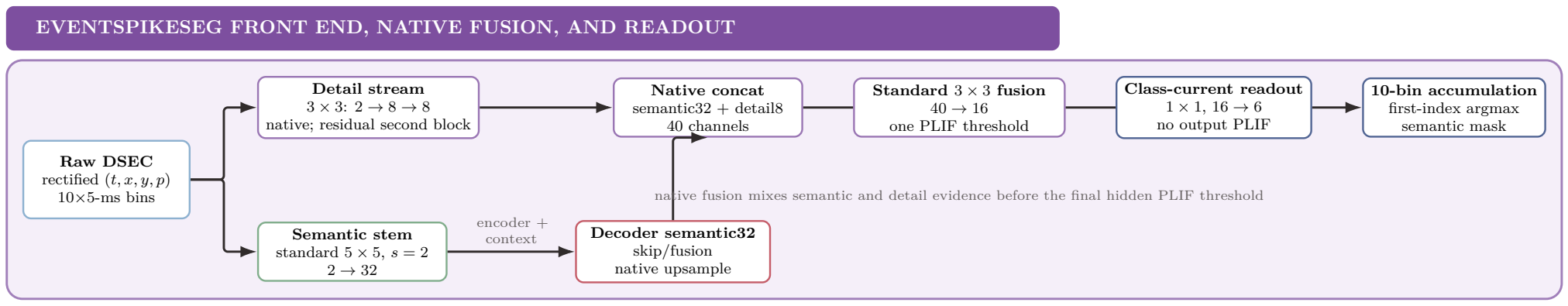
\begin{figure}[!t]
\centering
\begin{adjustbox}{max width=0.97\textwidth}
\begin{tikzpicture}[node distance=7mm]
  \node[precard, minimum width=30mm, minimum height=14mm] (raw) {\textbf{Raw DSEC}\\rectified $(t,x,y,p)$\\10$\times$5-ms bins};
  \node[bridgecard, right=12mm of raw, yshift=13mm, minimum width=34mm] (detail) {\textbf{Detail stream}\\$3\times3$: $2\rightarrow8\rightarrow8$\\native; residual second block};
  \node[enccard, right=12mm of raw, yshift=-13mm, minimum width=34mm] (sem) {\textbf{Semantic stem}\\standard $5\times5$, $s=2$\\$2\rightarrow32$};
  \draw[flow] (raw.east) -- ++(6mm,0) |- (detail.west);
  \draw[flow] (raw.east) -- ++(6mm,0) |- (sem.west);
  \node[deccard, right=23mm of sem, minimum width=35mm] (d2) {\textbf{Decoder semantic32}\\skip/fusion\\native upsample};
  \draw[flow] (sem) -- node[note,above,text width=20mm] {encoder +\\context} (d2);
  \node[bridgecard, right=24mm of detail, minimum width=34mm] (cat) {\textbf{Native concat}\\semantic32 + detail8\\40 channels};
  \draw[flow] (d2.north) |- (cat.south);
  \draw[flow] (detail)--(cat);
  \node[bridgecard, right=9mm of cat, minimum width=38mm] (fuse) {\textbf{Standard $3\times3$ fusion}\\$40\rightarrow16$\\one PLIF threshold};
  \node[readcard, right=9mm of fuse, minimum width=34mm] (read) {\textbf{Class-current readout}\\$1\times1$, $16\rightarrow6$\\no output PLIF};
  \node[readcard, right=9mm of read, minimum width=32mm] (arg) {\textbf{10-bin accumulation}\\first-index argmax\\semantic mask};
  \draw[flow] (cat)--(fuse)--(read)--(arg);
  \node[note, below=8mm of fuse] (n) {native fusion mixes semantic and detail evidence before the final hidden PLIF threshold};
  \panelbg{(raw)(detail)(sem)(d2)(cat)(fuse)(read)(arg)(n)}{bridgeLight}{bridgeDark!72}{panelV}
  \panelheader{panelV}{bridgeDark}{EVENTSPIKESEG FRONT END, NATIVE FUSION, AND READOUT}{190mm}
\end{tikzpicture}
\end{adjustbox}
\caption{EventSpikeSeg native-input, fusion, and continuous-readout path. The figure emphasizes the operations most relevant to recurrent-state cost and preservation of weak spatial evidence. Diagram preparation used AI-assisted LaTeX/TikZ coding, reviewed by the authors.}
\label{fig:eventspikeseg_path}
\end{figure}

Fig.~\ref{fig:eventspikeseg_path} expands the accuracy- and state-sensitive native path. Raw DSEC events feed a narrow native detail stem and a learned stride-2 semantic stem; decoded semantic features are concatenated with native detail features, mixed by a standard $3\times3$ PLIF stage, and accumulated by the continuous semantic readout.

\subsection{Float Reference and Training Boundary}
The analog architecture uses the original best-validation float checkpoint from epoch 58. Quantization-aware training (QAT) and Q7.16 digital exports do not define the analog numerical contract. Float32 parameter storage likewise does not imply float32 circuit accuracy: acceptable voltage/charge errors depend on model sensitivity and physical operating conditions.

Training carries recurrent state chronologically and resets it at stream breaks. The scheduled loss transfers emphasis from pixel cross-entropy (CE) toward target-present-class macro CE, increases Dice and boundary supervision, and decreases auxiliary CE. The native detail stream preserves fine spatial evidence while the semantic hierarchy learns broader context. Surrogate gradients train the hidden spikes and PLIF dynamics jointly. The analog reference is the best-validation epoch of the original seed-17, 80-epoch float run; later QAT checkpoints are not used for the analog mapping.

\subsection{Capacity Reference and Numerical Boundaries}
EventSpikeSegXL was retained as a capacity comparison: 979,179 registered parameters, 968,831 inference parameters, 959,824 synaptic weights, and 44 PLIF actors. Its additional lower-resolution processing, projected bridges, richer decoder fusion, three-way context, 16-channel detail path, 24-channel refinement, and four-head spiking attention explain the larger inventory. Attention uses Q/K coincidence and V gating rather than dense softmax. XL has a separate graph and readout configuration, and is excluded from the present analog measurements.

The earlier digital workflow appended three epochs of range-matched QAT with FP32 master parameters and forward-grid fake quantization. Its archived formats were Q7.16 weights/thresholds, Q15.16 membrane/current, UQ1.23 retention, six-bit adaptation age, Q3.16 adaptation amplitude/strength, and Q5.16 duration. These details preserve the design history; the active analog implementation uses the original float parameters and does not require QAT.

\subsection{Hardware-Oriented Model Selection}
\label{sec:evolution}

EventSpikeSeg was selected through several controlled search stages rather than by scaling a single backbone. The intermediate candidates are summarized here by the features they tested and the design decisions they established.

The initial ten-model screen varied standard versus depthwise-separable processing, dilation, multiscale fusion, normalization, auxiliary supervision, and skip structures. The strongest configuration was the FPGA-oriented hybrid design. This stage established that compact depthwise/pointwise processing could remain competitive, modest dilation was useful, and parameter count alone was a poor predictor of segmentation quality.

A first top-five mechanism screen then explored learnable PLIF dynamics together with instantaneous and longer-term inhibition. The strongest configuration kept the PLIF/long-term dynamics without aggressive hidden gating, indicating that hard competition could damage deeper spike propagation. A second top-five mechanism screen isolated peer-membrane inhibition, winner-take-all behavior, and triangular threshold adaptation. The light peer-membrane configuration was strongest, so EventSpikeSeg retains peer inhibition only at the stem together with local triangular threshold adaptation and rejects hard hidden winner competition.

The structural top-five search compared compact width, wider multiscale processing, additional depth, dilation, and an early large-kernel stem. The compact dilated configuration was strongest, reinforcing the choice of moderate semantic widths instead of widening every stage. The final top-three accuracy-oriented search focused on multiscale evidence preservation; the dual-path detail configuration was strongest and motivated the persistent native-resolution detail stream used in EventSpikeSeg.

The final EventSpikeSeg configuration therefore uses native $440\times640$ DSEC input, a narrow detail stream alongside a 32/64/128 semantic hierarchy, spike-preserving late fusion, and continuous temporal class-current accumulation rather than a spiking output population. These choices increase immutable synaptic capacity where useful while reducing spatially replicated recurrent state and avoiding premature loss of weak semantic evidence.

The architecture-search measurements were obtained under earlier or reduced protocols and are not used as final DSEC accuracy results. Their role is to explain why the retained EventSpikeSeg features were selected.

The hardware consequence is a large spatial state inventory despite a small synaptic parameter count. The model contains \SynWeights{} deployed synaptic weights but \NumPLIF{} membrane identities. Parameter sharing across spatial positions does not eliminate the need to retain each recurrent state. This motivated banked analog storage and shared physical service circuitry instead of replicating a complete analog update core for each logical neuron.

\section{Cross-Layer System Mapping}
\label{sec:mapping}
\subsection{Actor Graph, Persistent State and Communication}
The executed model is a directed actor graph in which a source spike can request signed contributions at one or more destinations. Each actor retains a tensor of membrane variables across successive chronological bins; the persistent state is not equivalent to one register per trainable synaptic parameter. The deployed graph has 25 PLIF actor populations and 54 directed edges, with 143,824 synaptic weights but 30,976,000 individual membrane identities. Adaptation age and amplitude together contribute another 21,964,800 identities, while class-current accumulation contributes 1,689,600. The resulting total is 54,630,400 logical state identities (Table~\ref{tab:state_inventory}).

Architecturally, this separation leads to three different replication domains: (i) logical state, whose independent values must be retained, (ii) reusable parameter and synaptic contribution logic, and (iii) physical service circuits that can be multiplexed over stored states. Convolutional weight sharing compresses the parameter domain but leaves the state domain proportional to spatial resolution and channel count. Replacing one compute core per state with bank/page-shared service reduces peripheral replication, but incurs addressing, read/restore, settling, and writeback transactions. The work therefore reports state, service, and connectivity inventories separately, rather than equating model parameter count with neuromorphic hardware size.

\subsection{Chronological Execution and Service Requirements}
The software loader creates ten ordered 5-ms model updates per 50-ms semantic-label record. In each update, input events are accumulated in their corresponding temporal bin, hidden spikes propagate through the defined graph, and the PLIF dynamics evolve with persistent state; an empty incoming-event bin does not imply the state is frozen. The implementation must preserve the trained recurrence and the specified order of leak, threshold/reset, inhibition, and adaptation. Output is a six-class continuous current accumulated across model updates, not six additional spiking neuron populations.

A service-capacity calculation highlights why sharing requires scheduling analysis. If an architecture attempted to update all 30,976,000 membrane identities once in every 5-ms interval, it would demand at least
\begin{equation}
 R_{\mathrm{state,all}}=\frac{30{,}976{,}000}{0.005\,\mathrm{s}}=6.1952\times10^9\ \mathrm{updates/s},
\label{eq:conditional_service}
\end{equation}
\emph{before} event fanout, memory access, or inter-actor dependency costs. This is a conditional all-state-workload calculation, not a measured EventSpikeSeg transaction rate or a universal lower bound for alternative sparse schedules. Whether silent states can be represented more economically while preserving leak and adaptation is a separate algorithm--architecture equivalence problem. The original project sets a 5-ms model-time accumulation target for supplied event traffic; it does not establish that all graph layers or a final segmentation mask can complete within 5 ms on the research layout.

\subsection{Hardware/Software Boundary and Reproducibility}
The software reference is the original floating-point seed-17 checkpoint. This establishes graph connectivity, membrane update semantics, and the recorded segmentation scores, but not analog numerical accuracy. The circuit mapping exposes signed contribution magnitude and sign, operating phases, analog reference/programming ports, and digital event/destination control. The original trained weights are used in the calibrated behavioral forward experiment in Section~\ref{sec:forward}, with ideal external programming and decoding. Physical trained-weight transfer and autonomous full-network inference have not been demonstrated. A meaningful co-design validation therefore needs (a) a workload trace of actual event fanout and service transactions, (b) measured or simulated access/compute/commit delays including routed loads, (c) an explicit schedule satisfying dependencies, and (d) hardware-in-the-loop or circuit-calibrated accuracy across the validation split. The present forward experiment addresses circuit-calibrated fidelity on one record across the graph; validation-split coverage and a physically realizable service schedule remain open.

\begin{table}[t]
\centering\small
\caption{Workload/implementation correspondence and current evidence.}
\label{tab:cosystem}
\begin{tabularx}{\textwidth}{p{0.26\textwidth}X}
\toprule
System dimension & Documented implementation and evidence boundary \\
\midrule
Model execution & 25 PLIF actors, 54 directed edges; original float model and chronological DSEC validation. \\
Persistent storage & 30.976 million membranes, plus adaptation and class-current identities; native 64-cell storage banks in GDS. \\
Event communication & Digital packet/actor/destination controls, addressed analog contribution interfaces; bounded RTL/connectivity checks, not total workload throughput. \\
Neuron service & Shared restore, signed accumulation, leak, decision/reset, and commit modules; nominal module/tile transients, not all-state timing. \\
Dense readout & Six continuous class-current outputs; native readout module evidence, no full-hardware mIoU. \\
Physical result & Hierarchical GF180 research atlas; no compact die area, silicon measurement, or system-level PPA. \\
\bottomrule
\end{tabularx}
\end{table}

\section{Banked Mixed-Signal Architecture}
\label{sec:asic}
\subsection{3T--MIM Membrane Storage and Shared PLIF Service}
The implementation retains the 25-actor model graph and its spatial identities. Resident cells use three-transistor access/sensing circuitry and a native metal--insulator--metal (MIM) capacitor; the characterized 8$\times$8 bank contains 64 stored states with shared column bias. The nominal membrane capacitor is 50 fF with a 5$\times$5-$\mu$m plate. Hierarchical page access connects banks to shared read, compute, and writeback circuits. Sharing removes repeated peripheral circuitry, while introducing access and settling costs; it does not merge independently stored membrane values.

The local operating sequence comprises restoring a selected resident state to the shared compute path, applying retention and signed contributions, and committing the updated state to storage. The resident capacitor then holds the state between accesses. Threshold and reset are PLIF service functions; the connected accumulation/hold experiment isolates update, commit, and retention with output-decision/reset stimuli disabled. Accordingly, the bank, PLIF core, and connected tile have distinct characterization boundaries.

Analog service modules implement PLIF update, signed contribution injection, threshold/reset, adaptation, inhibition, normalization, and class-current readout. Digital control macros handle packets, actor ingress, and destination dispatch. These control bits address analog states; they do not make the resident membrane a fixed-point register. The final hierarchy includes 54,630,400 logical analog state identities, of which 30,976,000 are membranes. Table~\ref{tab:state_inventory} separates the state categories; auxiliary compute capacitors and parameter/reference storage are outside this logical count.

\begin{table}[t]
\centering\small
\caption{Logical analog state inventory of the mapped inference model.}
\label{tab:state_inventory}
\begin{tabular}{lr}
\toprule
State category & Identities \\
\midrule
PLIF membrane & 30,976,000 \\
Adaptation age & 10,982,400 \\
Adaptation amplitude & 10,982,400 \\
Continuous class current & 1,689,600 \\
\midrule
Total & 54,630,400 \\
\bottomrule
\end{tabular}
\end{table}

\subsection{Signed Contributions and Model Time}
A presynaptic event triggers a signed contribution. Weight controls its magnitude and sign, not the presynaptic firing frequency. Pulse-duration/current injection and later switched-capacitor studies address the same integration requirement: deliver the intended signed charge while controlling state-dependent error, idle droop, and access overhead. The final research interfaces permit externally supplied magnitude/sign, references, clocks, and service phases. The layout therefore represents the architecture and native circuitry without claiming that all trained weights have been programmed or that autonomous end-to-end inference has been demonstrated.

For a nominal linear storage capacitor, the intended injection relation is
\begin{equation}
\Delta V_m=\frac{Q_{\mathrm{inj}}}{C_m},\qquad
Q_{\mathrm{inj}}=\int_{t_0}^{t_1}i_{\mathrm{inj}}(t)\,dt.
\end{equation}
A model-to-voltage scale must relate the accumulated model contribution to $\Delta V_m$. The trained floating-point representation does not determine that scale or the usable dynamic range. The reported pulse experiments characterize the native circuit under specified stimuli; they do not establish an accurate transfer function for all trained weights.

Contributions are accumulated before thresholding at the model update boundary. Learned retention is applied once per update; thresholding, subtractive reset, inhibition, and adaptation must follow the software ordering. The 5-ms requirement concerns supplied-event accumulation onto membranes, not completion of a segmentation output in 5 ms. A small connected test does not establish a population-wide deadline.

\subsection{Physical Graph and Interfaces}
The delivered hierarchical assembly retains 25 actor regions and verifies 947 native address nets reaching 1,622 endpoints. It records 2,075 actor boundary escapes, 118 class-current boundary escapes, and 725 verified native operation pads. Six categorical upper supply-grid components and four class supply/reference joins are documented. These counts describe checked connectivity scopes, rather than whole-chip timing or electrical signoff.

\section{Implementation Methodology and Verification Scope}
\label{sec:verification}

\begin{figure}[!t]
\centering
\includegraphics[height=87mm,width=0.97\linewidth,keepaspectratio]{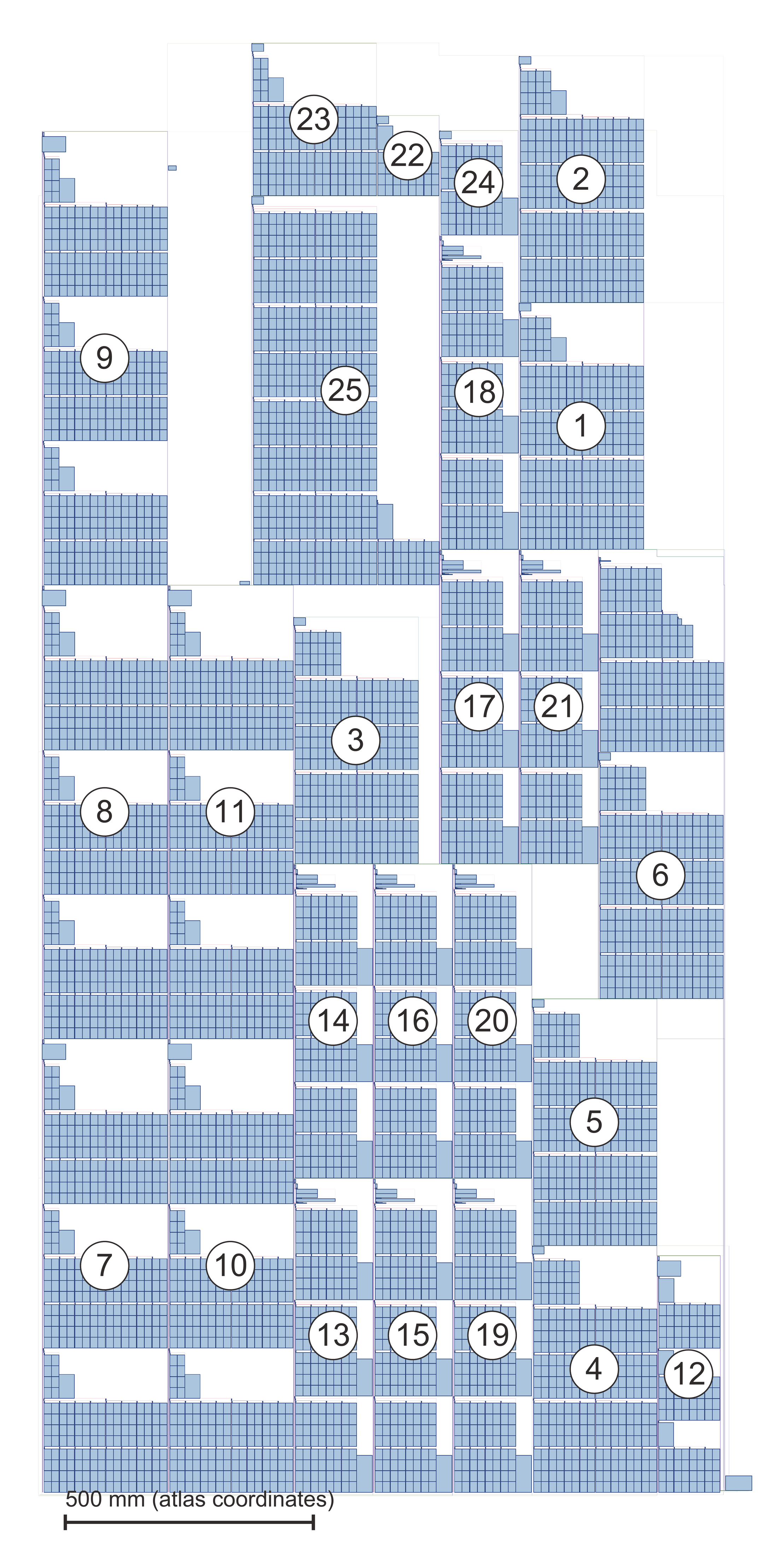}
\caption{Complete layout hierarchy overview derived from the delivered GDS, with unchanged relative placement. Markers 1--25 locate the individual regions in Supplementary Material S1. Filled blue regions include simplified small-cell extents; this is not a fully resolved transistor view. The atlas spans approximately $1{,}438.2\times2{,}929.0$ mm in database coordinates. That intentionally spacious assembly envelope is not a compact die-area result. Native component geometry is shown separately in Fig.~\ref{fig:gds_detail}. Rendering scripts were developed with AI assistance; the view is derived from GDS data.}
\label{fig:gds_full}
\end{figure}
Representative consumption is reported once per distinct physical element or shared module. The evidence combines native transistor transients, reused complete supply-current traces, native settled/DC samples, and explicitly labeled OpenROAD/OpenSTA library estimates. These methods provide different evidence boundaries; all reported circuit results are simulation-based, and no fabricated-chip measurements are included. The full connected 1,024-state page lacks a completed supply transient; its coverage is through frontend/leaf partitions.

\subsection{Characterization Conditions and Scope}
The 17 retained per-instance consumption jobs use a 3.3-V supply, 27$^\circ$C, and the typical GF180 device corner. Their supplied waveforms, loads, and measurement windows are retained with each result. Ideal independent fixtures drive programming and control ports; the energy boundary excludes their generators unless terminal imports are explicitly recorded. These nominal transient results are separated from the reused traces, settled samples, and library-based estimates described above.

The connected signed-update experiment uses the extracted native 64-state compute tile with 579 MOS devices, eight resistors, 172 native MIM instances, and 5,337 extracted capacitance branches. It uses a typical-corner 3.3-V, 27$^\circ$C setup, a 5-$\mu$s positive pulse starting at 97 $\mu$s and a 3-$\mu$s negative pulse starting at 110 $\mu$s. Statistical process/mismatch switches and output-decision/reset stimuli are disabled. The trace covers 5.23 ms and contains 2,360 adaptive time samples. Local routing capacitance is included; global atlas wiring RC and the routed duration-controller chain are outside this experiment. Consequently, the result characterizes signed accumulation, commit, and hold at one operating point.

For an operating window $[t_0,t_1]$, supply energy and mean power are
\begin{align}
 E_{\mathrm{VDD}}&=\int_{t_0}^{t_1}V_{\mathrm{DD}}(t)I_{\mathrm{in}}(t)\,dt,\\
 \overline P_{\mathrm{VDD}}&=E_{\mathrm{VDD}}/(t_1-t_0).
\end{align}
Here positive current enters the measured module. Independent programming, reference, load, and driver ports can supply additional energy; positive terminal imports are reported separately where measured. Energy can include stored capacitor energy and standing bias, so it must not automatically be interpreted as irreversible dissipation or incremental event energy. Shared-core, submodule, bank, and tile measurements overlap and must not be summed into a chip total.

\section{Evaluation: Software and Hardware Evidence}
\label{sec:results}
Table~\ref{tab:analog_consumption} reports per-module consumption, and Fig.~\ref{fig:state_transient} reports the connected extracted-tile voltage response. Together they provide electrical simulation evidence in addition to the physical views. The consumption inventory mixes explicitly distinguished analysis methods; it is not a full-chip post-layout power simulation.

\subsection{Delivered Layout and Native GDS Figures}
\begin{figure}[!t]
\centering
\begin{minipage}[t]{0.32\textwidth}\centering\vspace{0pt}
\includegraphics[height=36mm,width=\linewidth,keepaspectratio]{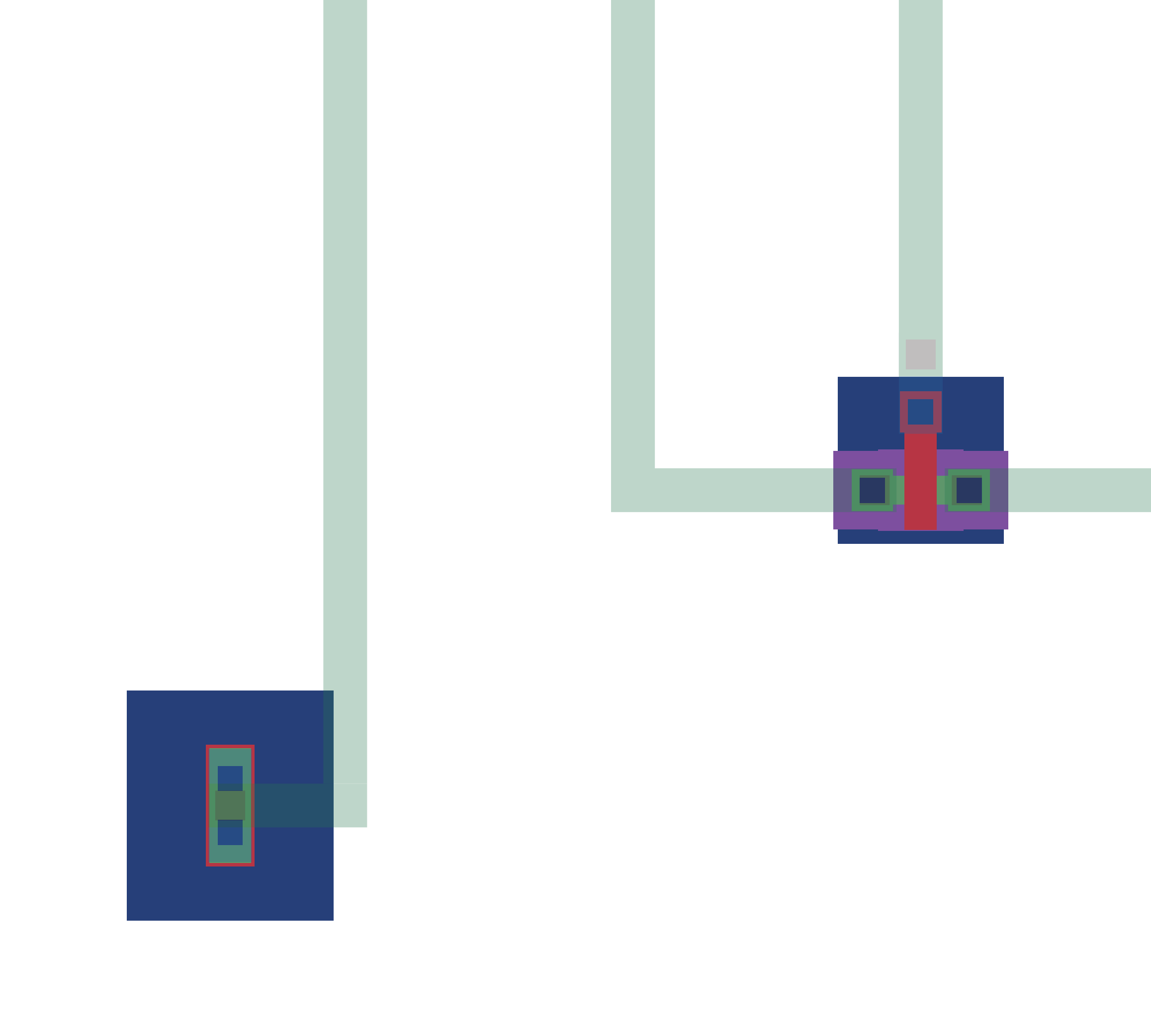}\\
{\small State/threshold devices: $10\times9\,\mu$m}\end{minipage}\hfill
\begin{minipage}[t]{0.32\textwidth}\centering\vspace{0pt}
\includegraphics[height=36mm,width=\linewidth,keepaspectratio]{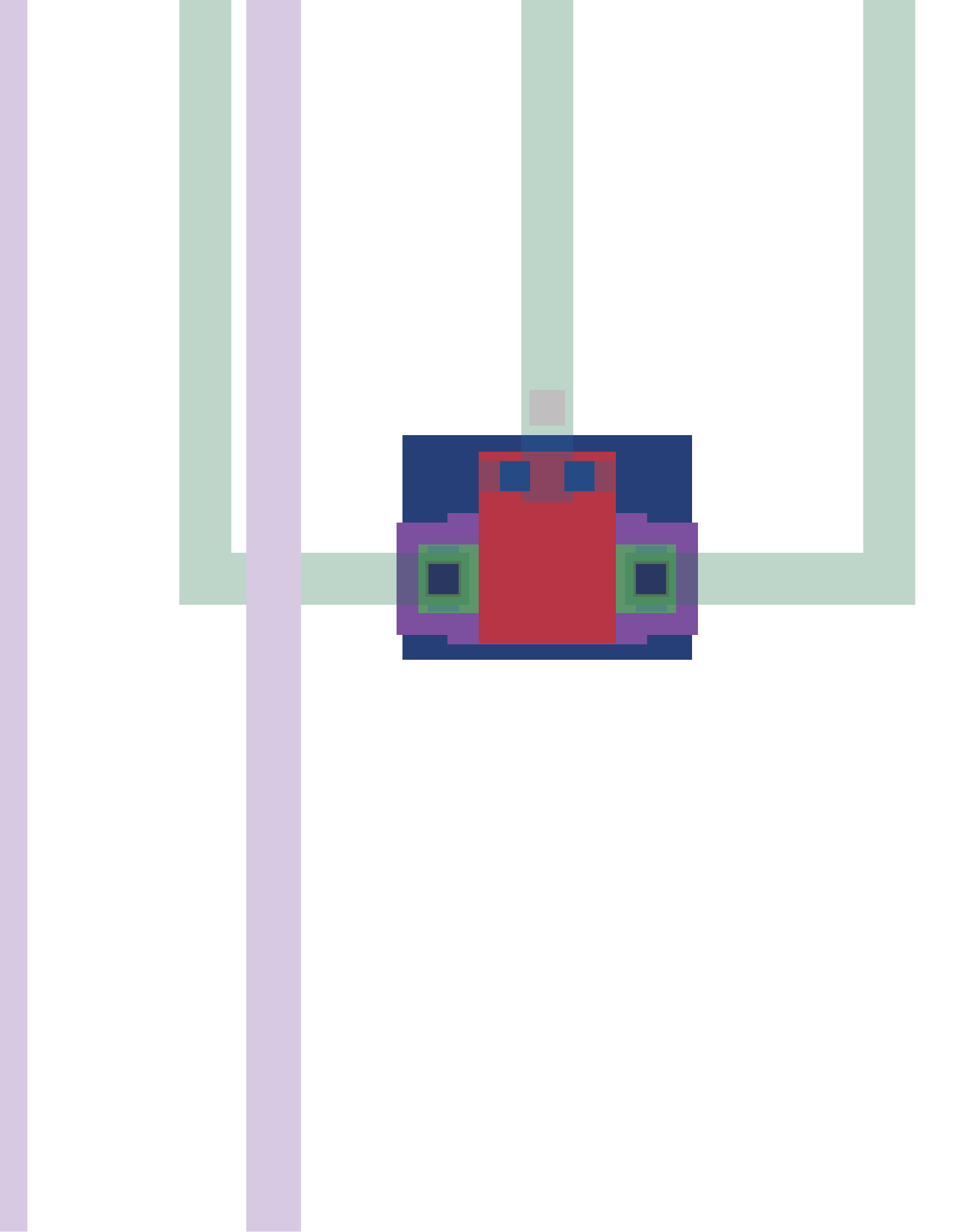}\\
{\small Reset device: $7\times9\,\mu$m}\end{minipage}\hfill
\begin{minipage}[t]{0.32\textwidth}\centering\vspace{0pt}
\includegraphics[height=36mm,width=\linewidth,keepaspectratio]{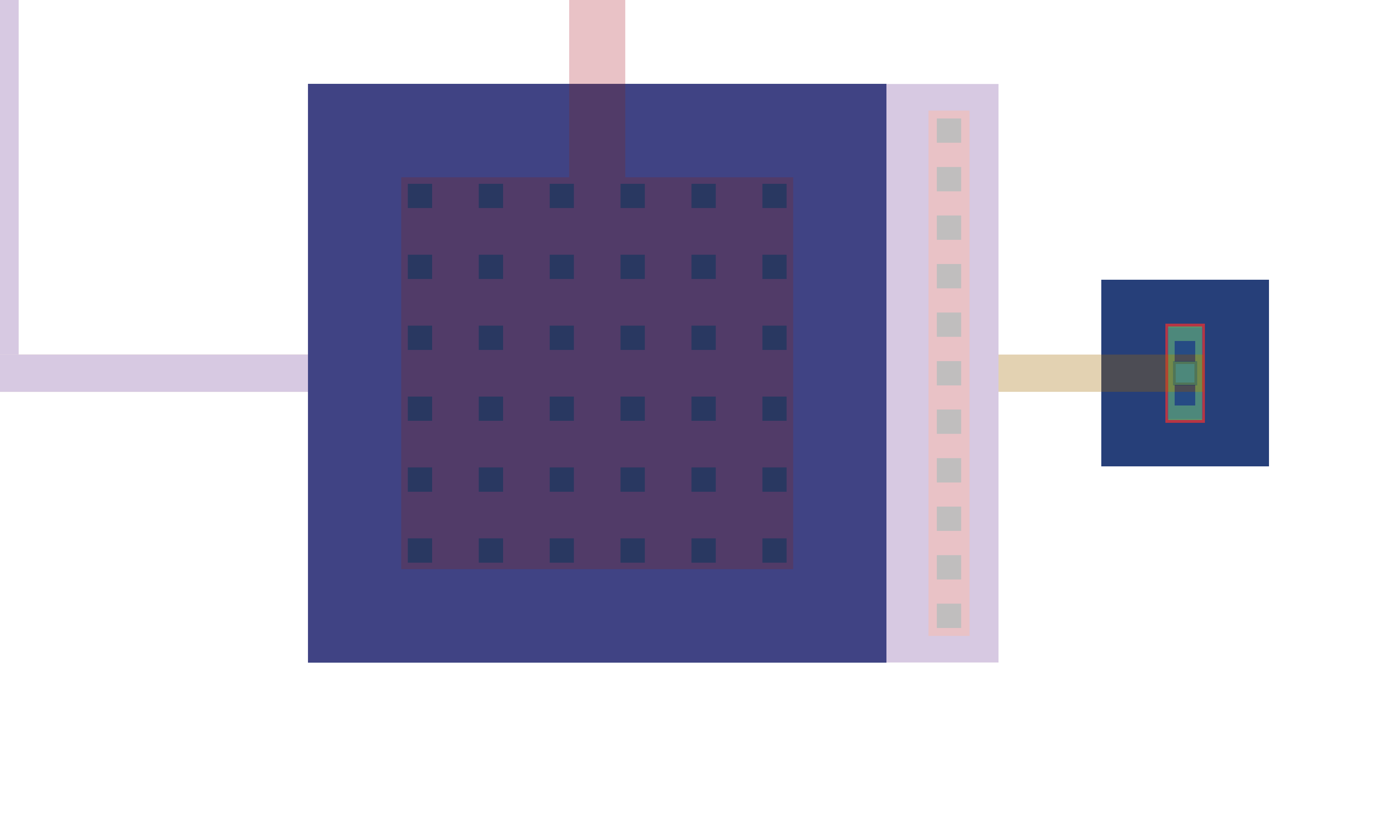}\\
{\small Decision storage region: $15\times9\,\mu$m}\end{minipage}\\[2mm]
{\small (a) PLIF device sections.}\\[3mm]
\begin{minipage}[t]{0.44\textwidth}\centering\vspace{0pt}
\includegraphics[width=0.88\linewidth]{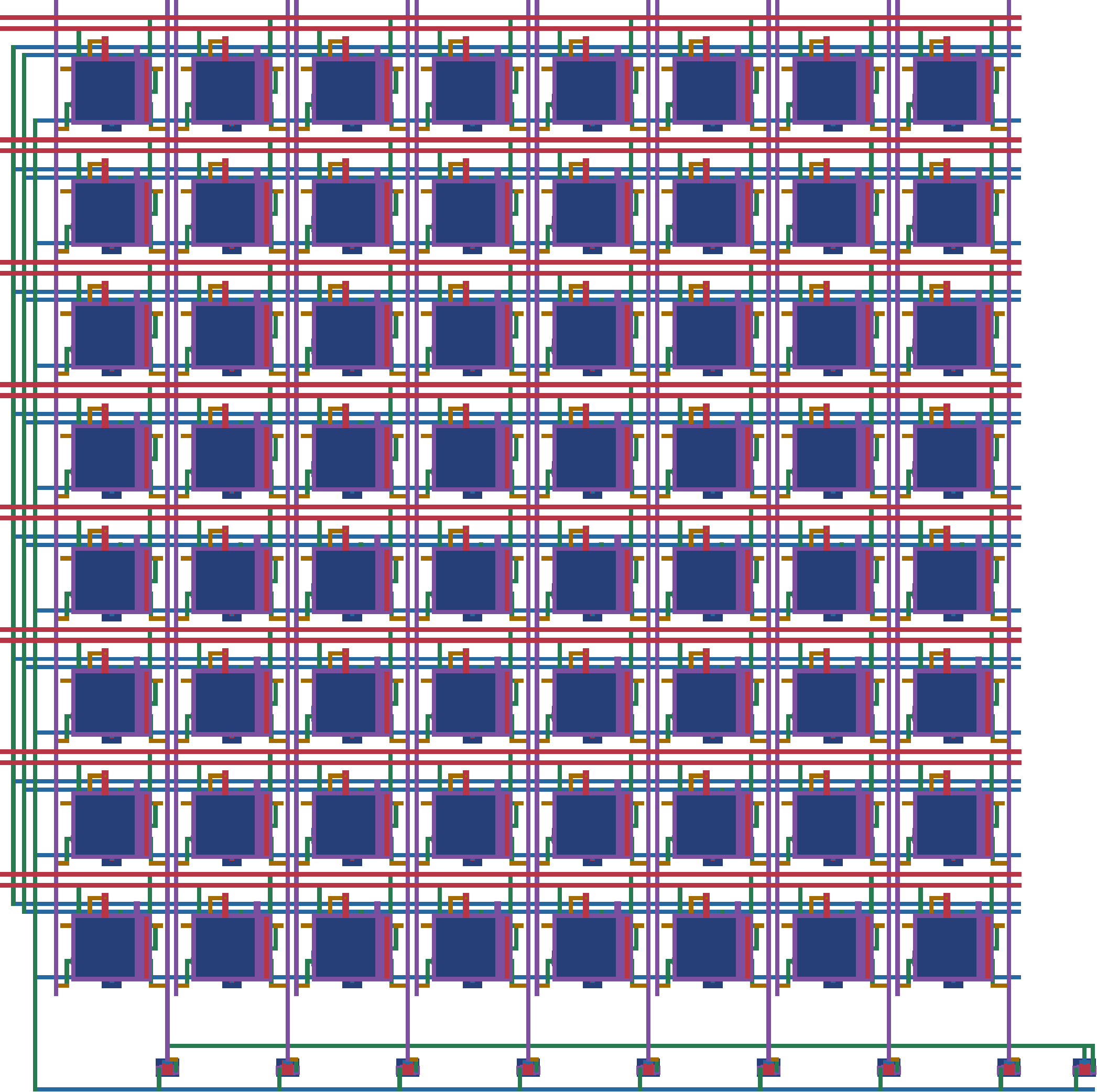}\\
{\small (b) Native 64-state bank: $100.32\times99.90\,\mu$m.}
\end{minipage}\hfill
\begin{minipage}[t]{0.52\textwidth}\centering\vspace{0pt}
\includegraphics[height=34mm,width=\linewidth,keepaspectratio]{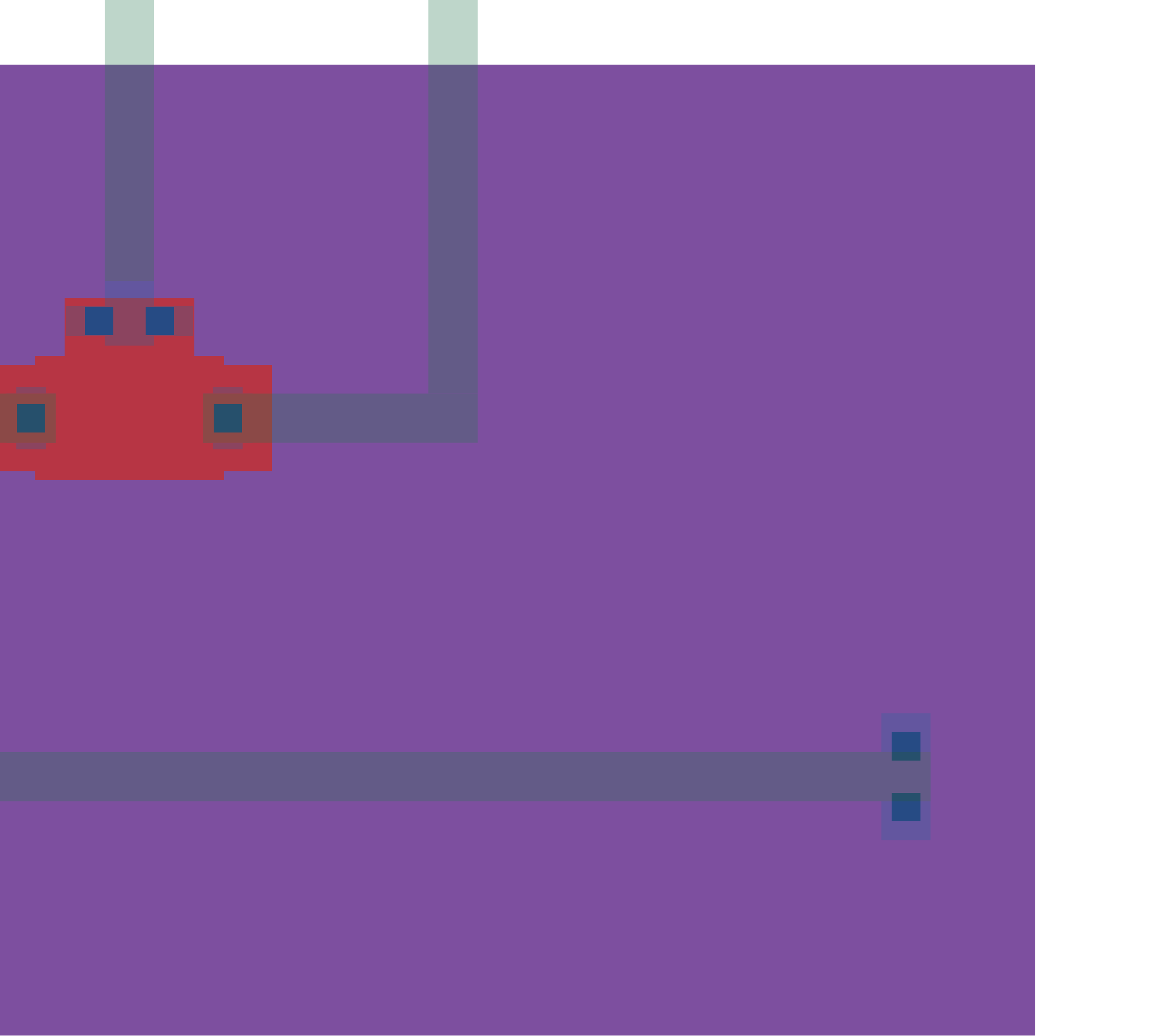}\\
{\small Source device: $9\times8\,\mu$m window}\\[2mm]
\includegraphics[width=54mm]{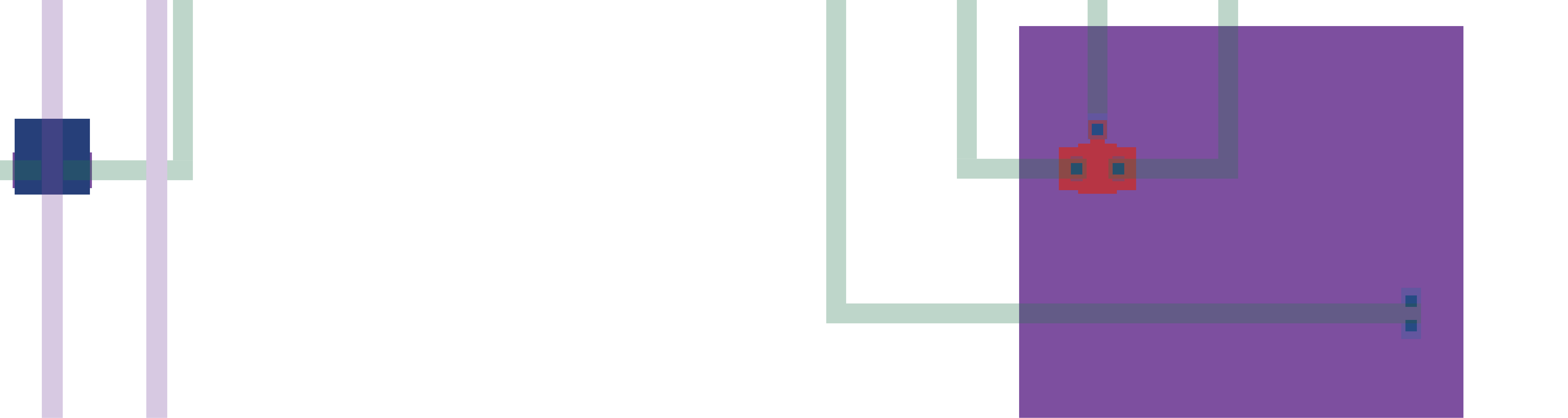}\\
{\small Transmission gate: $30\times8\,\mu$m window}\\[2mm]
{\small (c) Signed injector devices.}
\end{minipage}
\caption{Native GF180 layout views. The source GDS is unchanged. Views use independent scales; dimensions describe the displayed windows except for (b), which shows the full native 64-state bank. Panel (a) uses the delivered assembly; (b)--(c) use native component GDS. Rendering scripts were developed with AI assistance; the view is derived from GDS data.}
\label{fig:gds_detail}
\end{figure}

The hierarchical assembly includes the full mapped state inventory and the implemented module families. It is a spatially expanded research layout (an assembly atlas), whose envelope is not a compact die-area metric. Fig.~\ref{fig:gds_full} shows the complete placement overview. Fig.~\ref{fig:gds_detail} shows native mask geometry rendered by KLayout, including a PLIF cell selected inside the delivered full GDS. The storage-bank and injector views come from retained native component GDS files. The overview preserves the GDS coordinates while representing small hierarchy instances by extents. Numbered regions link the overview to Supplementary Material S1 (individual Figs. S1--S25), which provides an individual GDS-derived vector view of every actor region. The white-background palette improves print contrast; it does not alter placement or routing.

\subsection{Simulated Module Power and Operation Energy}
Supplementary Material S2 (Circuit Simulation Evidence) provides the complete 74-row consumption table, per-module reports, and the retained testbenches, simulator logs and waveforms. Its linked results index identifies the evidence for each operating point, including the PLIF core, resident membrane cell and signed injector; transistor simulations and digital-library estimates are explicitly distinguished.

The completed inventory covers all 22 required element/module families and contains 74 rows: 59 per-instance native-transient rows, six reused transient-supply rows, three native DC points, three settled transient samples, and three routed-library estimates. Table~\ref{tab:analog_consumption} lists representative operating windows. The PLIF core consumes 41.71 $\mu$W while biased; its 45-$\mu$s initialized beta/decision/reset window consumes 1.872 nJ with the injector parked. That window is not a complete weighted synaptic update. The single-cell read uses an external 1-$\mu$A column-load pulse and a 10-fF read load; the native device partition excludes bank-wire extraction.

\begin{table}[!htbp]
\centering\footnotesize
\setlength{\tabcolsep}{3pt}
\caption{Representative native module consumption; windows and boundaries are not interchangeable.}
\label{tab:analog_consumption}
\begin{tabularx}{\textwidth}{@{}>{\raggedright\arraybackslash}p{2.45cm}>{\raggedright\arraybackslash}p{1.35cm}>{\raggedleft\arraybackslash}p{1.82cm}>{\raggedleft\arraybackslash}p{2.02cm}>{\raggedright\arraybackslash}X@{}}
\toprule
Element / operation & Window & \shortstack[l]{Mean VDD\\power\\($\mu$W)} & \shortstack[l]{VDD energy\\(pJ)} & Characterization boundary \\
\midrule
PLIF core, biased hold & 5 ms & 41.711 & 208,555 & One physical core, including standing bias. \\
PLIF update & 45 $\mu$s & 41.603 & 1,872.15 & Injector parked; not full weighted update. \\
One resident cell, write & 1.2 $\mu$s & $1.197\times10^{-4}$ & $1.436\times10^{-4}$ & Positive terminal imports: 0.04123 pJ. \\
One resident cell, read & 1.4 $\mu$s & 2.382 & 3.3348 & Positive terminal imports: 3.33785 pJ. \\
Signed injector & 6 $\mu$s & 0.14845 & 0.89068 & Pulse/recovery; independent phase generation excluded. \\
Resident adaptation & 200 $\mu$s & 233.270 & 46,654.1 & Five age advances and two spikes. \\
Six-class evaluation & 100 $\mu$s & 330.599 & 33,059.9 & Readout evaluation only. \\
Shared bank-pair reader & 70 $\mu$s & 23.913 & 1,673.94 & Reused nominal supply trace. \\
Page reader frontend & 40 $\mu$s & 28.219 & 1,128.78 & Partition; not full 1,024-state page. \\
\bottomrule
\end{tabularx}
\end{table}

The 64-channel formatted normalization block has settled supply samples of 6.180--6.290 mW; the 128-channel unformatted reducer has samples of 3.453--3.674 mW. These are operating-point samples, not event energies, and their numerical errors remain documented. Routed-library default-activity estimates are 5.258 mW for the packet fabric, 75.950 mW for actor ingress, and 216.102 mW for destination dispatch. They retain the flow's activity/clock assumptions and are not measured full-chip or per-event consumption.

\subsection{Extracted-Tile Signed Update, Commit Error, and Retention}
The extracted-tile testbench, simulation report, simulator log and raw waveform supporting Fig.~\ref{fig:state_transient} are included in Supplementary Material S2.

A connected selected-state experiment produces positive and negative steps of +88.984 and $-254.945$ mV, respectively. The simulated commit error is $-19.979$ mV; other selected-row columns change by up to 14.377 mV. The selected-state drift over the following 5-ms hold is 0.531 mV. This establishes local signed-update and hold behavior while exposing commit and coupling errors. It does not establish calibrated weight transfer, noise robustness, or a full-population 5-ms service guarantee. Fig.~\ref{fig:state_transient} plots the retained binary simulation trace directly. The compute-node voltage and committed storage-node voltage are distinct circuit quantities; their absolute levels should not be treated as calibrated equivalents.

\begin{figure}[!t]
\centering
\includegraphics[width=0.98\textwidth]{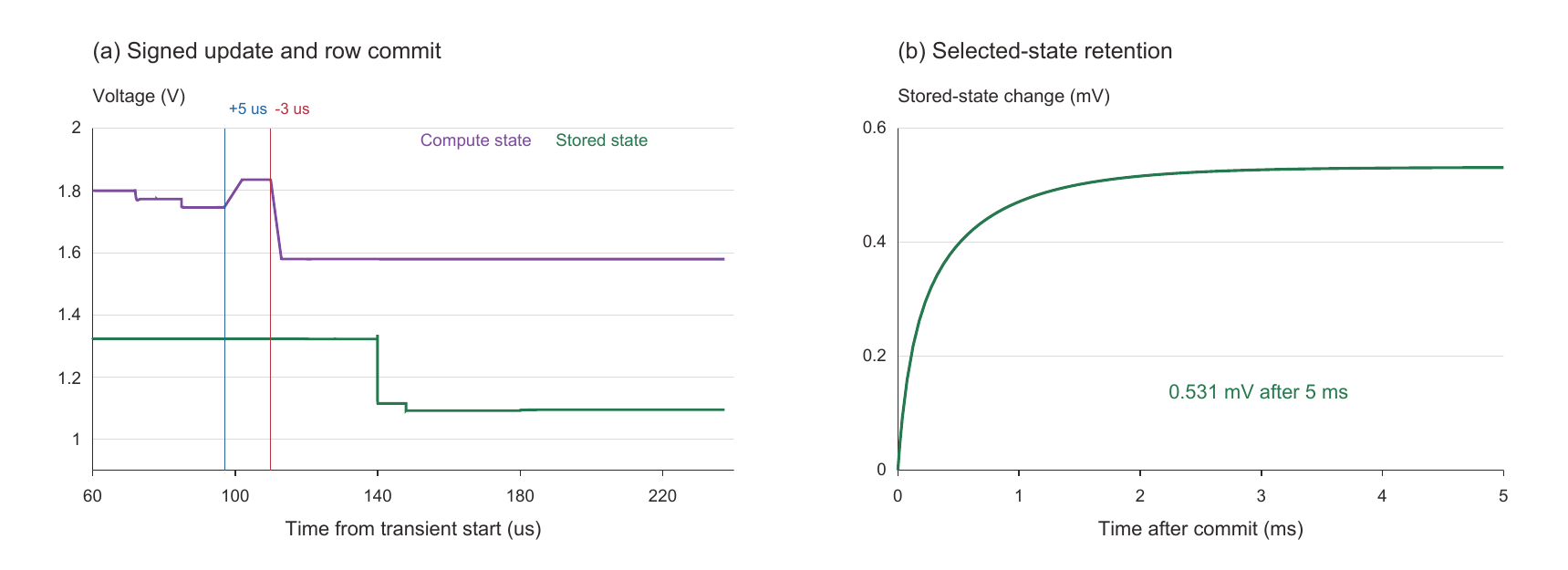}
\caption{Saved native-tile transient at 3.3 V, 27$^\circ$C, typical corner. (a) Shared compute state and selected resident state during signed pulses and row commit; pulse-start markers identify the externally driven positive and negative events. Restore is enabled at 70--80 $\mu$s and the beta phase at 85--85.5 $\mu$s. (b) Change in the selected stored voltage relative to its value at 230 $\mu$s, over the subsequent 5-ms hold. These plots reuse the original simulation data; no resimulation or weight fitting is applied. Plotting scripts were developed with AI assistance; plotted values are from the retained simulator output.}
\label{fig:state_transient}
\end{figure}

\subsection{Software Accuracy and Evidence Separation}
Software segmentation accuracy belongs to the selected PyTorch checkpoint and dataset protocol. It is not a measurement of the physical implementation. The retained epoch58 float-reference metadata records a best validation mean IoU of 0.2797627 (27.976\%). This is a stored checkpoint result under the recorded six-class DSEC protocol, not a new evaluation or a measurement of the ASIC. The matched circuit-calibrated behavioral result below is reported separately from this stored validation score and from physical hardware accuracy.

\subsection{SPICE-Calibrated Full-Frame Forward Simulation}
\label{sec:forward}
We completed a detailed behavioral forward pass for one $440\times640$ DSEC record (Zurich City 02a, frame 1), using the original floating-point epoch-58 checkpoint. All 25 actor convolutions were executed over ten chronological 5-ms event bins in two five-bin chunks, carrying neural state between chunks. The calculation uses a causal, per-pulse surrogate calibrated against native SPICE arithmetic, combined with PyTorch neural operators and a separately characterized nominal 5-fF storage variant. Integer event counts are encoded in each tap magnitude; it is not a transistor replay for every individual camera event. The reported result uses the detailed pulse-history model, not an endpoint approximation.

Storage errors are interpolated from native 64-cell SPICE simulations of fresh programming and two successive 5-ms reads. Separate curves represent event writes and inactive-state refresh. The simulation assumes additional Gaussian storage noise of 1 mV RMS (seed 17); this is an imposed sensitivity condition, not measured circuit noise. Encoded voltages span approximately 0.919--2.115 V, with no storage clipping or refresh-domain violations in this record. Repeated refresh extrapolates the two observed cycles; per-cell mismatch and physical refresh scheduling are not modeled.

External output compensation uses pulse count, native output voltage and programmed channel scale, without using the ideal convolution sum or segmentation labels to decode. On nine held-out native SPICE readout cases, compensation reduces RMS discrepancy from 6.016 to 0.729 mV, with a corrected maximum of 1.651 mV; one case lies outside the calibration grid. The pulse programmer, channel encoding, baseline subtraction and count/voltage decoder remain ideal external functions, rather than demonstrated physical circuits.

\begin{table}[t]
\centering
\caption{Matched one-record forward results on six-class DSEC. Both executions use the same floating-point checkpoint, input and target. These are single-record metrics, not validation-set or measured-chip accuracy.}
\label{tab:forward}
\begin{tabular}{lrr}
\toprule
Execution & mIoU (\%) & Pixel accuracy (\%) \\
\midrule
PyTorch reference & 26.22 & 64.17 \\
SPICE-calibrated behavioral model & 24.26 & 61.30 \\
Difference (percentage points) & $-1.96$ & $-2.86$ \\
\bottomrule
\end{tabular}
\end{table}

Table~\ref{tab:forward} establishes a completed model-level forward result under these assumptions. The CPU run took 890.98 s (14.85 min), with 2.42 GiB peak process memory; these are simulation resource costs, not accelerator latency or memory requirements. The single-record 26.22\% reference must not be confused with the checkpoint's stored 27.976\% validation mIoU. Predictions, targets, confusion matrices, matching reference, visual comparison, calibration tables and executed source snapshots are supplied as Supplementary Material S3 (Model Forward Simulation).

The empirical terminal/readout working tolerance is 3 mV on tested histories: observed long-history and readout discrepancies are 2.122 and 2.032 mV, respectively. The earlier 1-mV terminal target was not met. Among 309,760,000 arithmetic output values, 23,880 require history extrapolation and 637,083 require decoder extrapolation; the 3-mV tolerance does not apply outside the tested domain. Two native-source numerical checks also failed their original 0.01-mV precision target, with observed setting sensitivities of 0.488 and 0.166 mV. Thus the forward result extends the module evidence to one complete network input, while retaining numerical-source uncertainty, ideal interfaces and assumed noise. It does not establish accuracy across all 234 validation records, direct whole-GDS transistor inference, physical noise qualification, or a real-time service deadline.

\section{Discussion}
The dominant storage cost follows spatial state count. With the original assembly's nominal $5\times5$-$\mu$m MIM plate, one plate per membrane corresponds to 774.4 mm$^2$ of plate area alone. The retained native 64-state bank occupies a $100.32\times99.90$-$\mu$m bounding box, or 10,021.968 $\mu$m$^2$ per bank and 156.593 $\mu$m$^2$ per stored state after amortizing that bank's access, bias, and routing. Replicating this bank footprint for 30,976,000 membranes gives a geometric reference of 4,850.633 mm$^2$. This calculation excludes adaptation/class-current storage, higher-level service circuits, and top-level placement; it is not the final assembled union area or a compact-die estimate. It explains why capacitor plate area, bank footprint, and the much larger atlas envelope must be reported separately. Small synaptic parameter count does not imply small spatial recurrent-state hardware.

A separate compact bank now implements 64 nominal 5-fF cells, nine read-bias MOS devices and eight routed complementary row drivers, with 473 MOS devices in an $86.2\times87.9$-$\mu$m footprint. Its retained full DRC, unique LVS and GDS-readback checks pass; native 64-cell/two-refresh and charge/timestep checks support the storage model used in Section~\ref{sec:forward}. Its 7,576.98-$\mu$m$^2$ footprint is approximately 24.4\% smaller than the original bank, despite a tenfold reduction in nominal capacitance. Access, bias and routing therefore remain substantial. At an assumed effective density of 2 fF/$\mu$m$^2$, the plate-only estimate is 77.44 mm$^2$ for all membranes, whereas replicating the compact bank footprint would occupy approximately 3,667.3 mm$^2$ before higher-level circuits. Neither estimate is a compact die area. The original whole-model assembly remains unchanged; this new bank is a separately verified component, not a completed replacement throughout the hierarchy.

Sharing also creates a service-time tradeoff. The characterization windows include settling and bias costs, and the current interfaces can be externally driven. Global interconnect parasitics, supply-voltage drop and electromigration, full-chip timing, population service scheduling, and total chip power remain unqualified. The full 1,024-state connected-page transient and analog precision across noise, mismatch, and operating conditions remain open. These limits define the next engineering stages after the delivered architecture layout and representative characterization.

Historical selected-record software noise screens further show that analog accuracy cannot be inferred from a storage datatype. Those experiments are retained as additional research evidence, with their voltage scaling and limited record coverage explicit. The present contribution combines physical mapping, simulated module evidence and one calibrated full-frame behavioral inference. Broader accuracy evaluation, physical interface implementation and silicon area optimization require additional work.

\subsection{Architectural Interpretation and Comparison Limits}
The storage figures indicate that the dominant design challenge is not the raw count of learned weights but the scheduling and physical retention of a dense, spatially recurrent state graph. Moving computation into a shared service lane transfers complexity from replicated analog arithmetic to control traffic, bank-access contention, and voltage-error management. The retained nominal module results quantify some ingredients of that tradeoff but do not yet measure whether sharing is beneficial at application level. In contrast to fabricated co-processors such as Darwin \cite{ma2017darwin} or communication-aware SNN co-design \cite{xiao2025codesign}, the present implementation cannot support a quantitative energy, speed, area, or accuracy superiority claim.

For an application-level evaluation, the next experiments should use recorded DSEC event distributions to derive per-actor request arrival rates, compute worst-case and typical page-service occupancy, and measure deadline misses for explicitly stated scheduling and bank configurations. An independent sparse-state equivalence test would establish the cost of updating inactive neurons, while post-layout routed loads and broader calibrated analog replay would extend the one-record fidelity result to representative workload coverage. These are targeted validation requirements, not results supplied by the current experiments.

\section{Conclusion}
This work maps EventSpikeSeg to a banked GF180 mixed-signal research architecture, exposing constraints masked by its approximately 150k parameters. The 25-actor, 54-edge model entails 30,976,000 resident membrane states and 54,630,400 total logical state identities; the native 64-state banks and shared service modules separate storage replication from computation replication. Verified local connectivity and the hierarchical GDS demonstrate a physical research assembly. Nominal module simulations, including a 41.711-$\mu$W biased PLIF core and a connected signed-update/retention experiment, provide bounded implementation evidence, while the recorded 27.976\% six-class validation mIoU remains software-only. The detailed SPICE-calibrated forward pass achieves 24.26\% mIoU versus 26.22\% for the matched PyTorch frame across all 25 actors and ten event bins. This result uses a separately verified 5-fF bank variant, assumed 1-mV RMS storage noise and external count/voltage compensation. Broader accuracy coverage, physical encoding/decoding, feasible service scheduling and compact full-network implementation remain unresolved. The results should therefore be read as a resource-explicit architectural case study rather than a validated real-time neuromorphic segmentation accelerator.

\section*{Funding}
This research received no specific grant from funding agencies in the public, commercial, or not-for-profit sectors.

\section*{Declaration of Competing Interest}
The authors declare no known competing financial interests or personal relationships that could have appeared to influence the work reported in this paper.

\section*{Data Availability}
Supplementary Material S1 comprises 25 individual actor-region layout views (Figs. S1--S25). Supplementary Material S2 contains the circuit simulation evidence. Supplementary Material S3 supplies the completed model forward simulation, matched PyTorch reference and compact 5-fF bank evidence. All three are supplied in the accompanying supplementary archive, with the layout views retained as individual vector PDFs. It includes a linked results index, the 74-row module-consumption table, numerical reports, retained testbenches, simulator logs, and waveforms. These artifacts support the reported operating points but do not constitute a verified one-command reproduction environment.

The project repository for source code and documentation is \url{https://github.com/st-amirr/EventSpikeSeg}. The generated circuit evidence described above is supplied as Supplementary Material S2; a separate public repository deposit of these generated data has not yet been established. The experiments used only the publicly available DSEC dataset, including its DSEC-Semantic labels \cite{gehrig2021dsec,sun2022ess}. The dataset is available from \url{https://dsec.ifi.uzh.ch/dsec-datasets/download/}, with semantic-label information at \url{https://dsec.ifi.uzh.ch/dsec-semantic/}.

\section*{CRediT authorship contribution statement}
Amir Mohammad Khadem Hosseini: conceptualization, methodology, software, validation, formal analysis, investigation, data curation, visualization, original draft, review and editing, and project administration. Sattar Mirzakuchaki: supervision.

\section*{Declaration of generative AI and AI-assisted technologies in the manuscript preparation process}
During this work, the authors used OpenAI tools, including Codex, and xAI Grok to assist code generation and debugging, exploration and combination of model features, development of the final network model, and scripting and execution support for circuit layout, routing, and simulations. OpenAI tools were used primarily for these development tasks and also assisted manuscript organization and editing. OpenAI image generation assisted an earlier conceptual illustration; the model diagram in Fig.~\ref{fig:base_arch} is rendered from editable TikZ source. Physical-layout views were rendered from GDS, and transient plots were produced from retained simulation data using scripts. The authors reviewed and edited the assisted work and take full responsibility for the content of this publication.



\end{document}